\documentclass[conference]{IEEEtran}
\makeatletter
\def\ps@headings{%
\def\@oddhead{\mbox{}\scriptsize\rightmark \hfil \thepage}%
\def\@evenhead{\scriptsize\thepage \hfil \leftmark\mbox{}}%
\def\@oddfoot{}%
\def\@evenfoot{}}
\makeatother
\usepackage[cmex10]{amsmath}
\usepackage{array}
\usepackage{url}
\usepackage{makecell}
\usepackage{graphicx}
\graphicspath{{./figures/}}

\begin{document}
%
\title{\emph{PowerTracer}: Tracing requests in multi-tier services to save cluster power consumption}



%
\author{\IEEEauthorblockN{Lin Yuan\IEEEauthorrefmark{1},
Jianfeng Zhan\IEEEauthorrefmark{1},
Bo Sang\IEEEauthorrefmark{1},
Lei Wang\IEEEauthorrefmark{1} and
Haining Wang\IEEEauthorrefmark{2}}
\IEEEauthorblockA{\IEEEauthorrefmark{1}Institute of Computing Technology, Chinese Academy of Sciences, Beijing, China, 100190}
\IEEEauthorblockA{\IEEEauthorrefmark{2} Department of Computer Science, College of William and Mary  Williamsburg, VA 23187 }
}


\maketitle

\begin{abstract}
As energy proportional computing gradually extends the success of DVFS (Dynamic voltage and frequency scaling) to the entire system, DVFS control algorithms will play a key role in reducing server clusters' power consumption.
The focus of this paper is to provide accurate cluster-level DVFS control for power saving in a server cluster.  To achieve this goal, we propose a request tracing approach that online classifies the major causal path patterns of a multi-tier service\footnote{Triggered by an individual request, a causal path is a sequence of component activities with causal relations. \emph{Major} causal path patterns represent repeatedly executed causal paths that account for significant fractions.} and monitors their performance data of each tier as a guide for accurate DVFS control.
The request tracing approach significantly decreases the time cost of performance profiling experiments that aim to establish the empirical performance model. Moreover, it decreases the controller complexity so that we can introduce a much simpler feedback controller, which only relies on the single-node DVFS modulation at a time as opposed to varying multiple CPU frequencies simultaneously. Based on the request tracing approach,  we present a hybrid DVFS control system that combines an empirical performance model for fast modulation at different load levels and  a simpler feedback controller for adaption. We implement a prototype of the proposed system, called \emph{PowerTracer}, and conduct extensive experiments on a 3-tier platform. Our experimental results show that PowerTracer outperforms its peer in terms of power saving.
\end{abstract}

\section{Introduction}
%
%
%
%

In data centers, most of services adopt the multi-tier architecture, and services are replicated or distributed on a cluster of servers\cite{23Urgaonkar2005}. Server cluster operating costs arising from increasing energy consumption of  non energy-proportional server hardware \cite{25Barroso2007} \cite{non_proportional} have prompted the investigation of cluster-level power management.
Current solutions that save cluster power consumption can
be classified into three main categories: (1) DVFS (Dynamic
voltage and frequency scaling), (2) dynamic cluster
reconfiguration by consolidating services through request distribution \cite{19Horvath2007} \cite{request_distribution_2003},
and (3) server consolidation through moving services to virtual
machines \cite{power_nap} \cite{co_co} \cite{PARTIC} \cite{euro_sys_2009}.

On one hand, reconfiguration approaches can greatly reduce cluster power consumption by consolidating services through request distribution \cite{19Horvath2007} or moving services to virtual machines\cite{power_nap} on a subset of physical nodes \cite{power_nap} and turn off the remaining ones. However, as claimed by D. Meisner et al \cite{power_nap}, reconfiguration alone, cannot close the gap of resource demands between peak and average workloads, and server clusters still require sufficient capacity for peak resource demand. On the other hand, as energy proportional computing \cite{25Barroso2007} gradually extends the success of DVFS to the entire system \cite{power_nap}, accurate DVFS control algorithms are essential to reducing cluster power consumption. Therefore, the motivation of this work is how to leverage commercial DVFS technologies, which exemplifies the energy-proportional concept \cite{power_nap},  on physical nodes to save cluster power consumption.

In this paper, we propose a novel request tracing approach \cite{26Barham2004}\cite{14Zhang2009} for accurate cluster-level DVFS control. Through instrumentation, which can be performed on operating system kernel, middleware or application levels, our request tracing tool can obtain  major \emph{causal path patterns} (in short, patterns) in serving different requests and capture server-side latency, especially service time of each tier in different patterns. Thus, we can fully understand the role of each component played in serving different requests in terms of \emph{service time percentage}.
The advantage of the request tracing approach in DVFS control lies in two-fold: first, it decreases the time cost of performance profiling experiments that aim to obtain the performance model; second, it decreases the controller complexity. Exploiting major patterns, we propose a simpler feedback controller, which only relies on the single-node DVFS modulation at a time as opposed to varying multiple CPU frequencies in a simultaneous manner. On a basis of this,  we present a hybrid DVFS control algorithm that combines an empirical performance model for fast modulation at different load levels and a simpler controller for adaption.

We implement a prototype of the proposed system, called \emph{PowerTracer}, and
conduct extensive experiments on a 3-tier platform. We use two kinds of workload of a RUBiS (Rice University Bidding System) web application \cite{15rubis} to validate the efficacy of PowerTracer in terms of three metrics: total system power savings compared to the baseline, request deadline miss ratio, and average server-side latency.  In particular, we investigate the effects of the server-side latency threshold and the number of main patterns upon server cluster power consumption and other performance metrics.
Our experimental results show that PowerTracer outperforms its peer \cite{19Horvath2007} in terms of power saving.

The remainder of this paper is organized as follows. Section \ref{background} presents the background of this work.
Section \ref{model_and_system} details the system architecture, performance model, and DVS controller of \emph{PowerTracer}. Section \ref{system_architecture_implementation} describes the system implementation. Section \ref{evaluation} presents the experimental results. Section \ref{related_work} surveys related work. Finally, Section \ref{conclusion} draws a conclusion.

\section{Background}\label{background}
In this section, we highlight the background information of {\em PowerTracer} with respect to request tracing, performance profiling, and control theory.

\begin{figure}[hbtp]
  \centering
  \includegraphics[scale=0.40]{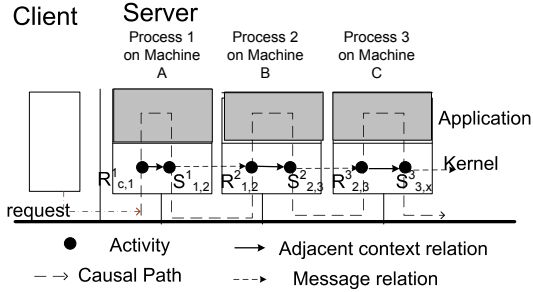}
  \caption{Activities with causal relations in the kernel.}
  \label{request_observation}
\end{figure}

\textbf{Request tracing}. We employ the concept of request tracing from our previous work \cite{14Zhang2009, scalable_tracing}. In Fig. \ref{request_observation}, we observe that a request causes a series of \emph{interaction activities} in the OS kernel, e.g. sending or receiving messages. Those activities happen under specific contexts (processes or kernel threads) of different components. We record an activity of sending a message as $S^{i}_{i,j}$, which indicates a process $i$ sends a message to a process $j$. We record an activity of receiving a message as $R^{j}_{i,j}$, which indicates a process $j$ receives a message from a process $i$. Our concerned activity types include: \emph{BEGIN}, \emph{END}, \emph{SEND}, and \emph{RECEIVE}. SEND and RECEIVE activities are those of sending and receiving messages. A BEGIN activity marks the start of servicing a new request, while an END activity marks the end of servicing a request.

When an individual request is serviced, a series of activities have causal relations or happened-before relationships, and constitute \emph{a causal path}. For example in Fig. \ref{request_observation}, the activity sequence \{$R^{1}_{c,1}, S^{1}_{1,2}, R^{2}_{1,2}, S^{2}_{2,3}, R^{3}_{2,3}, S^{3}_{3,x}$\} constitutes a causal path. For each individual request, there is a causal path.  For a request, the
{\em server-side latency} can be defined as the time difference between the time stamp of BEGIN activity and END activity in its corresponding causal path. The service time of each tier can be defined as the accumulated processing time between SEND activities and RECEIVE activities. For each tier, its role in serving a request can be measured in terms of \emph{service time percentage}, which is the ratio of
service time of the tier to the server-side latency.

After reconstructing those activity logs into causal paths, we classify those causal paths into different \emph{patterns}. Then, we present main patterns to represent
repeatedly executed causal paths, which account for significant fractions. For each pattern, we can obtain the average server-side latency and average service time of each tier. In addition, through observing the number of BEGIN activities, which marks the beginning of serving new requests, we can derive the current load level of the services.

\textbf{Performance profiling}. On a basis of the approach used in \cite{SLA_decomposion}, our approach aims to create detailed performance profiles of multi-tier services with different DVFS modulations under varying workloads. A tier profile captures each tier's performance characteristics as a function of DVFS modulation and varying loads. In order to profile a tier, we deploy a test environment and vary the DVFS setting of a node, on which each tier is deployed. We then apply a variety of loads and collect each tier's performance characteristics, independent of other tiers. After acquiring the measurements, we derive the appropriate functions mappings from DVFS modulation and load levels to each tier's performance metrics.

\textbf{Control theory}.
Similar to the control-theoretic terminology used in \cite{control_theory},
we refer to the server cluster and deployed multi-tier service being controlled as \emph{the target system}, which has a set of metrics of interest, referred to as \emph{measured output}, and a set of control knobs, referred to as \emph{control input}. The controller periodically (at each \emph{ control period}) adjusts the value of the control input such that the measured output (at each \emph{sampling period}) can match its desired value----referred to as \emph{reference input} specified by the system designer. We refer to the difference between the measured output and the reference input as \emph{control error}). The controller aims to maintain control error at zero, in spite of the disturbance in the system.

\section{Architecture and System Design}\label{model_and_system}
In this section, we first describe the architecture of PowerTracer. Then, we present
design details in online request tracing, performance profiling, and controller, respectively.

\subsection{PowerTracer Architecture}\label{powertracer_architecture}

\begin{figure}[hbtp]
  \centering
  \includegraphics[scale=0.40]{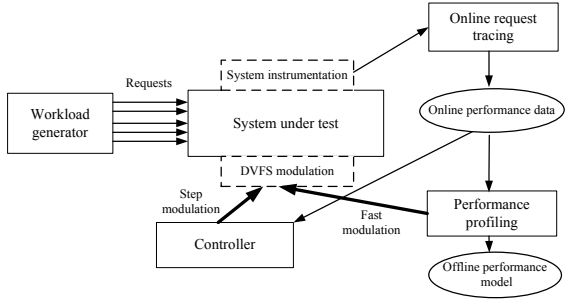}
  \caption{\emph{PowerTracer} system architecture}
  \label{system_architecture}
\end{figure}

As shown in Fig.\ref{system_architecture}, PowerTracer consists of three major modules: online request tracing, performance profiling, and controller.
The online request tracing module is used to collect online performance data of multi-tier services, including main patterns, their respective server-side latencies, service time of each tier, and current loads.

The performance profiling module aims to off-line establish an empirical performance model under different load levels. The model can capture the relationship between service performance and CPU frequency setting of each node. With the online performance data produced by the request tracing module, we can decide the dominated tier by checking service time percentage of each tier in different patterns. Then, we mainly observe the effect of DVFS modulation of the node, on which the dominated tier are deployed. In this way, we can ease the establishment of the performance model.

For different load levels, which can be provided by the request tracing module, the controller module quickly calculates the optimal setting of DVFS modulation according to the performance model created by the performance profiling module. For adaptation, the controller module only relies on the single-node DVFS modulation at a time as opposed to varying multiple CPU frequencies simultaneously.

\subsection{Online Request Tracing}\label{request_tracing_system}
Our previous work PreciseTracer \cite{14Zhang2009} \cite{scalable_tracing} has described how to design and implement an online request tracing system for multi-tier services of black boxes. Black boxes indicate that we do not need to have source code of application or middleware. In this paper, our contribution is how to exploit online request tracing for accurate DVFS control in multi-server clusters.

On a basis of PreciseTracer, we develop a new component called \emph{Analyzer}, which classifies a
large variety of causal paths into different patterns, then extracts online performance data for main patterns according to their fractions.

Our classification policy of causal paths addresses the following challenging problems: first, it is difficult to utilize individual causal path from massive request traces as the guide for DVFS modulation; second, causal paths are different and the overall statistics of performance information (e.g. the average server-side latency used in \cite{19Horvath2007}) of all paths hides the diversity of \emph{patterns}.
There are several ways of classifying causal paths. For example, in our previous work \cite{scalable_tracing}, we classify causal paths into different patterns according to their shapes. In this paper, we found that classifying causal paths according to \emph{the size of the first message sent by a client} performs well in producing patterns. Thus, we use a k-means cluster method \cite{22Hartigan1979} to classify causal paths into patterns based on the size of the first message sent by a client.

We use a 5-tuple \textless{}\textit{pattern ID}, \textit{the size of the first message send by a client}, \textit{the average server-side latency of pattern ID}, \textit{the average service time per tier}, \textit{the current load}\textgreater{} to represent the online performance data of a pattern in a multi-tier service. In our previous work \cite{scalable_tracing}, for RUBiS workload \cite{15rubis}, we observe that there are more than hundred patterns, but the top ten patterns take up a large fraction of paths, more than 88\%. Therefore, we single out the top $N$ patterns according to their fractions as the guide for building both the performance model and the controller module.

\subsection{Performance profiling}\label{performance_profiling}
The purpose of performance profiling is to create an empirical performance model, which is used for the fast modulation procedure. The power model can be represented by Equation (\ref{equation_1}) as follows.


\begin{equation}\label{equation_1}
\begin{cases}
D_{L,1}=\sum_{j=1}^{M}f_{L,1,j}(F_{j})+\gamma_{L,1}\\
\vdots\\
D_{L,N}=\sum_{j=1}^{M}f_{L,N,j}(F_{j})+\gamma_{L,N}\end{cases}\end{equation}

$N$ is the number of  {\em major} patterns singled out by our request tracing system. In our experiments in Section \ref{evaluation}, we will observe the effect of different number of main patterns on the power saving.  $M$ is the tiers of a multi-tier service, namely three in this paper.  \textit{D$_{L,i}$}$(i=1,2,...,N)$ is the average server-side latency of pattern $i$ when the current load is $L$.
$\gamma_{L,i}$($i$=1,...,$N$) is the network latency of pattern $i$ when the current load is $L$. Function \textit{f$_{L,i,j}$(F$_{j}$)} represents the average service time of each tier $j$ in pattern $i$ when the clock frequencies run with \textit{F$_{j}$}(j=1,...,M) and the current load is $L$.

For each load level, the function set $f$ is derived as follows: first, through our tracing system, we measure the service time of each tier while traversing all the four discrete CPU frequency values offered by each server; second, we derive functions \emph{ f$_{L,i,j}$} between the average service time of each tier in major patterns and the CPU frequencies of nodes by the normal quadratic polynomial fitting tool. The models fit well, and the fitting coefficient \emph{R$^{2}$\textgreater{}}97\%.

Through request tracing and performance profiling, we can ease the establishment of the empirical performance model. For each workload, e.g. given the transition table and the number of current clients, we create a function set $f$ offline. For a three-tier service deployed
on the testbed that offers four CPU frequency levels, if we consider 10 different load levels (from zero to the upper bound of the load), we need to do $10\times 4 \times 3$ times of experiments to gain the function set $f$. Since our request tracing tool can obtain the service time of each tier in major causal path patterns and decide the dominated tiers in terms of the service time percentage of each tier,  our system can effectively decrease the time cost of performance profiling experiments by mainly scaling the CPU frequencies of the tiers that dominate the server-side latency. For example, for RUBiS, the average service time percentage of the web server, the application server, and the database server are 0.11\%, 17.63\%, and 82.26\%, respectively. So, we consider the database server as the dominated tier. In this way, we can decrease the times of experiments to $10\times 4\times 1$.

\subsection{Controller design}\label{feedback_control_system}

In our feedback controller, the target system is the multi-tier server cluster. The control inputs are the new clock frequency vector. The measured outputs are the server-side latencies of the top main $N$ patterns, which are presented as a vector as well.   We define \textit{TH$_{i}$} (i=1,...,N) as the server-side latency threshold of pattern $i$. The reference inputs are the desired server-side latency {\em threshold zones} for the main patterns. $LP$ and $UP$ represent the lower and upper threshold factor. The controller module makes the measured outputs fall within the latency threshold zone by adjusting the values of the control inputs periodically.

The controller module consists of two main procedures: fast modulation and step modulation. During the fast modulation procedure, by tracing the current load, PowerTracer can decide the current load level and compute out the optimal clock frequency of each server with reference to the given server-side latency threshold zones of the main patterns according to Equation (\ref{equation_1}).

After the fast DVFS modulation, it is the step of modulation loop.
In our system, step modulation periods are composed of alternate \emph{sampling} periods and \emph{control} periods. We model our system as Equation \ref{equation_3}.
\begin{equation}\label{equation_3}
\overrightarrow{D(t+1)}=F(\overrightarrow{D(t)},\overrightarrow{f(t)})\end{equation}
$\overrightarrow{D(t)}$ is a vector that represents the average server-side latencies of each pattern in the $t_{th}$ sampling period. ${\overrightarrow{f(t)}}$ is a vector that represents the CPU frequency levels of the nodes in the $t_{th}$ period. $\overrightarrow{F(t)}$ is a vector that represents the transition functions of each pattern from states of the $t_{th}$ period to that of the $(t+1)_{th}$ period.

We design the step modulation procedure as follows.
In the $t_{th}$ sampling period, we use the request tracing module to obtain 5-tuple online performance data of the top $N$ patterns.
%
By comparing the measured server-side latencies of the top $N$ patterns with the latency threshold zones, the controller module modulates the DVFS setting based on the following procedure.

For pattern $i$, if $D_{i}(t)$ exceeds its upper threshold \textit{UP*TH$_{i}$}, the controller module chooses tier $j$ that has the maximum service time to step up its CPU frequency and record the new frequency value into the frequency vector. The frequency values of the other tiers remain intact. For any pattern $i$ ($i=1,..,N$) $D_{i}(t)$ fall below their lower threshold \textit{LP*TH$_{i}$}, the controller module chooses tier $j$ that has the minimum service time to step down its CPU frequency and record the new frequency into the frequency vector. The frequency values of the other tiers remain intact.

The approach that we take the measured server-side latencies of the top \textit{N} main patterns as the controlled variables, enables the majority of requests meet service-level agreement. A small percentage of requests is ignored so as to achieve the most power reduction. The step modulation procedure can be formally defined in Equation \ref{equation_4}.

\begin{equation}\label{equation_4}
\begin{cases}
f_{j}(t+1)=f_{j}(t)+1, & \exists i|(D_{i}(t)>UP*TH_{i})\wedge \\(ST_{i,j}(t) &is\_the\_maximum);\\
f_{j}(t+1)=f_{j}(t)-1, & \forall i|(D_{i}(t)<LP*TH_{i})\wedge \\(ST_{i,j}(t) &is\_the\_minimum); \\
f_{j}(t+1)=f_{j}(t), & otherwise\end{cases}\end{equation}

In Equation \ref{equation_4}, \emph{ST$_{i, j}(t)$} represents the service time of tier $j$ in pattern $i$ in the $t_{th}$ period. For $f_{j}$, {\em $+1$} indicates stepping up CPU frequency with one level, while {\em $-1$} indicates stepping down CPU frequency with one level.

\section{System Implementation}\label{system_architecture_implementation}

\begin{figure}[h]
\centering
\includegraphics[scale=0.55]{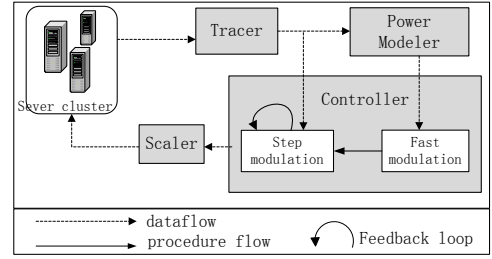}
\caption{The architecture diagram of PowerTracer.}
\label{system_implementation}
\end{figure}
As showin in Fig.\ref{system_implementation}, PowerTracer includes four major components: \emph{Tracer}, \emph{Power Modeler}, \emph{Controller} and \emph{Scaler}.

On a basis of our previous work \cite{scalable_tracing}, we implement the request tracing system described in Section \ref{request_tracing_system} and integrate it into PowerTracer as a module called Tracer. As shown in Fig.\ref{system_implementation}, Tracer reads all service logs and outputs 5-tuple performance data of the top $N$ patterns.

Through changing the DVFS modulation according to Section \ref{performance_profiling}, PowerModeler analyzes the 5-tuple performance data of the top $N$ patterns, and then outputs the power model in the file \emph{Pre\_model}.

For different load levels, Controller runs the fast modulation based on the Pre\_model file. Then, Controller runs step modulation loops, which are composed of alternate sampling and control periods. In the sampling period, Tracer is called to output 5-tuple performance data for the top $N$ patterns. In the control period, Controller makes decisions on changing clock frequencies as the approach presented in Section \ref{feedback_control_system}.

Scaler, which runs on each tier, is called by Controller to set clock frequencies. The clock frequency setting is invoked by setting frequency scaling governor of the Linux kernel and recording new frequency into scaling\_setspeed file.

\section{Evaluation}\label{evaluation}
We use a 3-tier web application RUBiS \cite{15rubis} to evaluate the efficiency of our approach in terms of system performance and power savings. RUBiS is a three-tier auction site prototype modeled after
eBay.com, developed by Rice University.
\subsection{Experimental Setup}\label{experimental_setup}
\begin{figure}[t]
\centering
\includegraphics[scale=0.55]{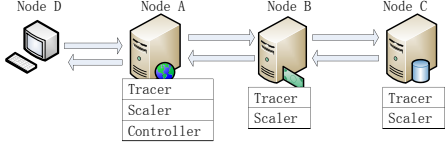}
\caption{The deployment diagram of 3-tier platform.}
\label{deployment}
\end{figure}

The testbed is a heterogeneous 4-node platform composed of Linux-OS blade servers,
which we name \emph{node A}, \emph{node B}, \emph{node C}, and \emph{node D}, respectively, as shown in Fig. \ref{deployment}. All four nodes are DVFS-capable. Nodes A and B both have 2 capable processors, which support the frequencies at 1.0, 1,8, 2.0 and 2.2GHz. Node C has 8 processors, which support the frequencies at 0.8, 1.1, 1.6 and 2.3GHz. Node D has 2 AMD Opteron (tm) Processors. We use QINGZHI 877X power analyzer to measure the power consumption. As shown in Fig. \ref{deployment}, the platform is deployed with 3-tier web servers as follows:
\begin{itemize}
\item  	Node A is deployed as the web server tier, which runs Apache ( server version 2.2.13).
\item  	Node B is deployed as the application server tier, which runs JBoss enterprise Java application (version 4.2).
\item  	Node C is deployed as the database server tier, which runs MySQL ( server version 5.0.45) database.
\item  	Node D runs RUBiS client emulator, which generates and sends requests to the Apache server.
\end{itemize}

\subsection{Workload}\label{workload}
We use the RUBiS client emulator to generate workload by adjusting the parameters, e.g. the number of clients and transition tables. In our experiments, we use two kinds
of transition tables, which emulate read\_only workload (browse\_only table) and read\_write mixed workload (transition table),respectively. For the read\_only workload, we set the number of clients to 100,
200, 300, 400, and 500, respectively. For the mixed workload,
in order to enhance experimental contrast effects, we set the number of clients  to 500, 600, 700, 800, and 900, respectively. Each workload includes three stages, of which we  set up ramp time, runtime session,  and down ramp time  as 10 seconds, 300 seconds, and 10 seconds, respectively.

As shown in Fig. \ref{deployment},
Tracer and Scaler are deployed on all three servers. For Controller, we only deploy it on Node A rather than the other two servers.
Note that the service delay at the web server tier (Node A) imposes the least impact on requests' performance.

In the experiments,  our prototype system is simplified in two aspects. First, we only modulate clock frequencies of the last two tiers (Jboss and Mysql servers) and leave the first tier running with its lowest clock frequency all the time. Second, we utilize a constant average value, which is gained by experiments, to represent the network latencies among the three tiers during the fast modulation procedure. The rationale behind the simplification is as follows: first, our experiments show that the service time of the first tier (Apache server), which is 10$^{-2}$ to 10$^{-1}$ milliseconds on average, contributes little to the server-side latency, which is 10 to 100 millisecond; second, current commercial network interface cards are not energy proportional, and we cannot change its setting as we modulate CPU frequencies to save power consumption.

\subsection{Experimental Results}\label{results}
We use three metrics to evaluate our system: total system power savings compared to the baseline, request deadline miss ratio, and average server-side latency. We set the clock frequency of all servers to the maximum as \emph{the baseline}.  Note that the power consumption under the baseline is not fixed for different load levels, and higher load will lead to higher power consumption even with the same clock frequency. \emph{We assume the server-side latency under the baseline is $\overrightarrow{SL}$}. For PowerTracer, $\overrightarrow{SL}$ is a vector representing the server-side latencies of the chosen main patterns.  For the request deadline miss ratio metric, we predefine a server-side latency deadline, which is under the SLA constraints.  We compare the results of PowerTracer with those of the two other algorithms: the SimpleDVS algorithm presented by Horvath et al. \cite{19Horvath2007} and the Ondemand governor offered by Linux kernel. The SimpleDVS algorithm takes CPU utilization as the indicator in determining which server's clock frequency should be scaled. The Ondemand governor, the most effective power management policy offered by Linux kernel,  keeps the CPU frequency low when not needed, and instantly jumps back to full power when required \cite{31ondemand}.  In particular, we implement a modified version from PowerTracer, called \emph{PowerTracer\_NP}, which chooses the average server-side latency as the measured output, instead of $N$ individual server-side latencies of the top $N$ patterns.

For all experiments, we set the sampling period as 1 second and the sampling interval as 5 seconds. The control period is 1 second. We set the upper latency threshold factor, \emph{UP}, and the lower latency threshold factor, LP, as 1.2 and 0.8, respectively. We trace the servers' performance statistics while setting the servers with different frequency values for different workload, and use the normal quadratic polynomial fitting to derive the performance model described in Section  \ref{performance_profiling}.

Note that the maximum power savings we can achieve in three nodes (node A, node B and Node C)  is 17.68\% when we set the clock frequencies of the three servers to the minimum without considering performance. We do not consider node D, because it is used to generate client requests.

\subsubsection{Read\_only workload}\label{read_only_workload}
\begin{figure}[h]
\centering
\includegraphics[scale=0.65]{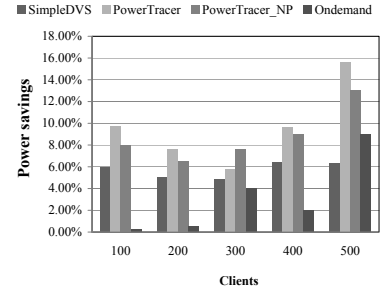}
\caption{A comparison of the total system power savings. The x-axis represents the number of clients set by RUBiS client emulator. The y-axis represents the total system power savings compared to the baseline. }
\label{browse_comparison_powersaving}
\end{figure}


\begin{figure}[h]
\centering
\includegraphics[scale=0.65]{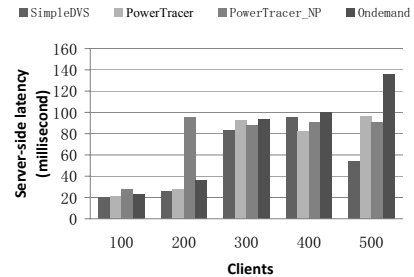}
\caption{A comparison of the average server-side latencies. The y-axis represents the average server-side latency of all requests.}
\label{browse_comparison_avgdelay}
\end{figure}


\begin{figure}[h]
\centering
\includegraphics[scale=0.65]{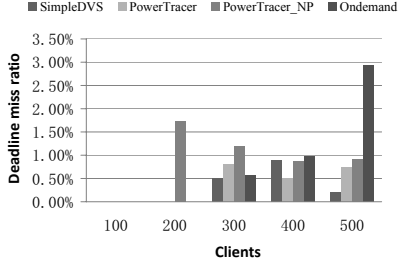}
\caption{A comparison of the request deadline miss ratios. The y-axis represents the deadline miss ratio of all requests and the blank represents zero. The deadline is set to 0.5 second for RUBiS read\_only workload.}
\label{browse_comparison_missratio}
\end{figure}


\begin{figure}[h]
\centering
\includegraphics[scale=0.60]{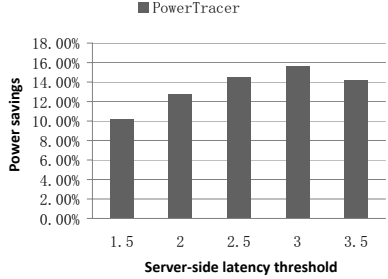}
\caption{The total system power savings of PowerTracer under five latency thresholds. The x-axis represents five latency threshold levels.}
\label{PowerTracer_Powersaving_5factor}
\end{figure}

Fig. \ref{browse_comparison_powersaving}, Fig. \ref{browse_comparison_avgdelay}, and Fig. \ref{browse_comparison_missratio} present the total system power savings, the average server-side latencies, and the request deadline miss ratios of the four algorithms, respectively, when the number of clients varies from 100 to 500. In this set of experiments, we set the server-side latency threshold as $3\times \overrightarrow{SL}$. For PowerTracer, \emph{the number of main patterns} is 5, which indicates that the top 5 patterns are used as the guide for DVFS control.  PowerTracer or PowerTracer\_NP gains the highest power reduction. The Ondemand governor gains the lowest power reduction except when the number of clients is 500. PowerTracer outperforms PowerTracer\_NP except when the number of clients is 300. When the number of clients is 500, PowerTracer gains the maximum power saving of 15.60\% compared to the baseline, which is about 147\% better than SimpleDVS\footnote{Note that in \cite{19Horvath2007} , when the SimpleDVS algorithm is performed on the three laptop computers with Mobile AMD Athlon XP DVS-capable processors, and the authors reported that the TPC-W service consumes as much as 30\% less energy compared to the same baseline as in our paper.}, and about 74\% better than the Ondemand governor in terms of power saving.  We also observe that Ondemand has poor performance in terms of  both the average server-side latency and the request deadline miss ratio when the number of clients is 500. This is because the Ondemand governor of each server scales CPU frequencies without any coordination.

\begin{figure}[t]
\centering
\includegraphics[scale=0.60]{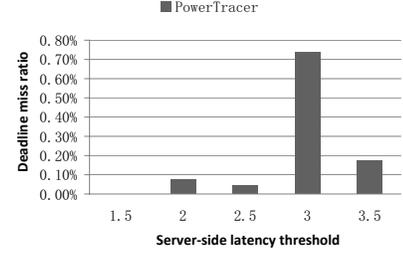}
\caption{The request deadline miss ratios of PowerTracer under five different latency thresholds.  The deadline is set to 0.5 second as above.}
\label{powertracer_missratio_5factor}
\end{figure}

\begin{figure}[t]
\centering
\includegraphics[scale=0.60]{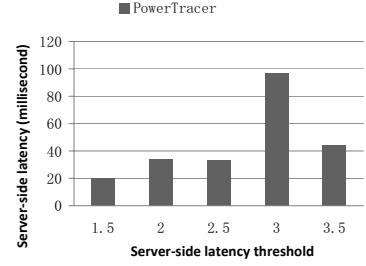}
\caption{The average server-side latencies of PowerTracer under five latency thresholds.}
\label{powertracer_avgdelay_5factor}
\end{figure}

From Fig.\ref{PowerTracer_Powersaving_5factor} to Fig.\ref{powertracer_avgdelay_5factor}, we demonstrate how the performance varies when the reference input in PowerTracer---the server-side latency threshold, varies. We set the number of clients to 500 and select five main patterns according to their fractions from the top seven patterns. For the top seven patterns, $\overrightarrow{SL}$ is (2.583734, 20.585, 25.88851, 63.86756, 74.38303, 63.06882, 66.01713) milliseconds. We set five latency thresholds as 1.5, 2, 2.5, 3, and $3.5 \times \overrightarrow{SL}$, respectively. From Fig.\ref{PowerTracer_Powersaving_5factor} to Fig.\ref{powertracer_avgdelay_5factor}, we can see that when the latency threshold  is $2.5 \times \overrightarrow{SL}$, PowerTracer is about 43\% better than that when the the latency threshold  is $1.5 \times \overrightarrow{SL}$ in terms of the power saving. At the same timer, under these two configurations, their server-side latencies and their request deadline miss ratios are close to each other. Therefore, choosing an appropriate latency threshold is critical to PowerTracer.

Fig.\ref{browse_powersavings_mainpatterns} \textasciitilde{} \ref{browse_avgdelay_mainpatterns} show the performance of PowerTracer when  a different number of main patterns is chosen out of the top seven major patterns. For these experiments, we set the number of clients to 500 and the latency threshold to $3 \times \overrightarrow{SL}$.
Fig.\ref{browse_powersavings_mainpatterns} \textasciitilde{} \ref{browse_avgdelay_mainpatterns} show that if we set the number of main patterns to one, PowerTracer gains the highest power saving, but also the highest server-side latency and the highest request deadline miss ratio.   Comparing the performance of one main pattern and that of five main patterns, we can see that choosing more patterns does not necessarily improve power savings and other metrics, and hence, we need to choose \emph{the number of main patterns} based on different workloads so as to achieve the optimal power saving and performance.


\begin{figure}[h]\label{browse_powersavings_mainpatterns}
\centering
\includegraphics[scale=0.60]{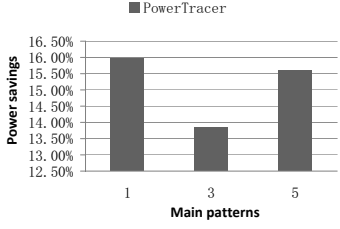}
\caption{A comparison of the total system power savings with different number of main patterns. The x-axis represents the chosen number of the main patterns.}
\label{browse_powersavings_mainpatterns}
\end{figure}

\begin{figure}[h]\label{browse_missratio_mainpatterns}
\centering
\includegraphics[scale=0.60]{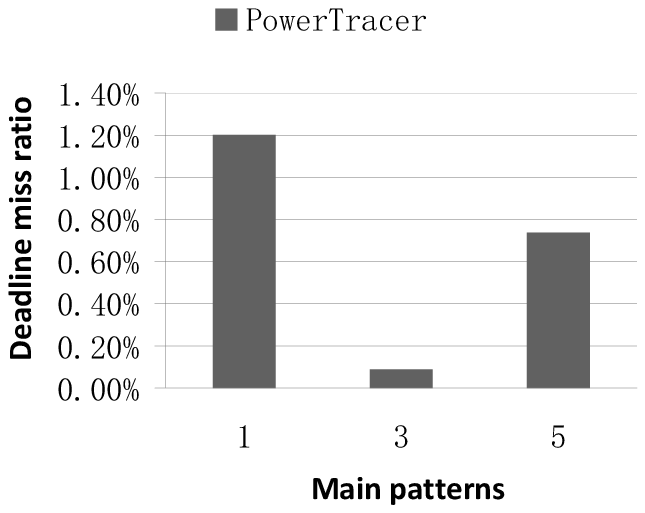}
\caption{A comparison of the request deadline miss ratios with different number of main patterns.}
\label{browse_missratio_mainpatterns}
\end{figure}

\begin{figure}[h]\label{browse_avgdelay_mainpatterns}
\centering
\includegraphics[scale=0.60]{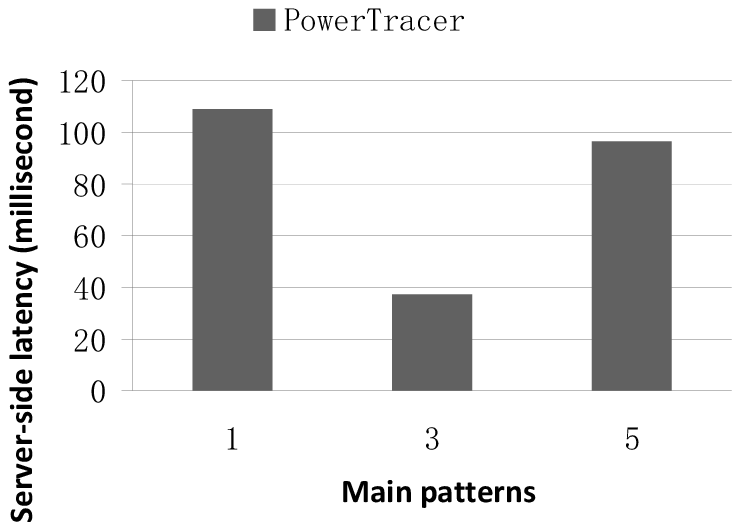}
\caption{A comparison of the average server-side latencies with different number of main patterns.}
\label{browse_avgdelay_mainpatterns}
\end{figure}

\begin{figure}[h]
\centering
\includegraphics[scale=0.65]{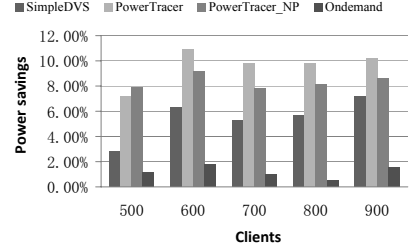}
\caption{A comparison of the total system power savings.}
\label{transition_comparison_powersaving}
\end{figure}


\subsubsection{Read\_write mixed workload}\label{read_write_mixed_workload}

\begin{figure}[h]
\centering
\includegraphics[scale=0.65]{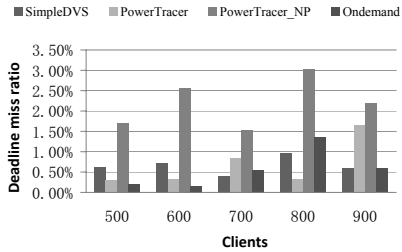}
\caption{A comparison of the request deadline miss ratios. The deadline target is set to 0.2 second for read\_write mixed workload.}
\label{transition_comparison_missratio}
\end{figure}


\begin{figure}[h]
\centering
\includegraphics[scale=0.65]{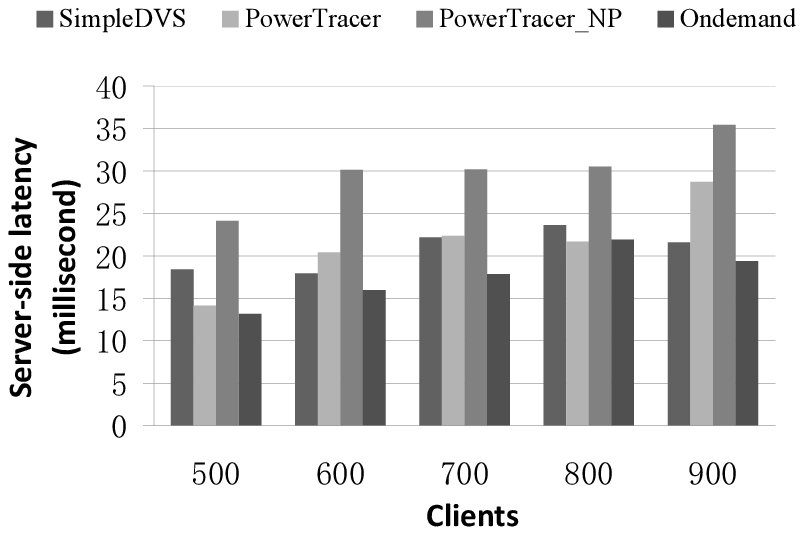}
\caption{A comparison of the average server-side latencies.}
\label{transition_comparison_avgdelay}
\end{figure}

Fig.\ref{transition_comparison_powersaving}, Fig.\ref{transition_comparison_missratio} and Fig. \ref{transition_comparison_avgdelay} present the total system power savings, the request deadline miss ratios, and  the average server-side latencies gained by the four algorithms, respectively, when the number of clients varies from 500 to 900. In these experiments, we set the number of the main patterns to 3 and the latency threshold  to $5 \times \overrightarrow{SL}$. We can see that PowerTracer or PowerTracer\_NP gains the highest power saving. PowerTracer outperforms PowerTracer\_NP except when the number of clients is 500. The Ondemand governor has the lowest power reduction. When the number of clients is 600, PowerTracer achieves the maximum power saving of 10.88\% compared to the baseline, which is about 74\% better than SimpleDVS and about 515\% better than the Ondemand governor in terms of power saving.

\begin{figure}[h]
\centering
\includegraphics[scale=0.60]{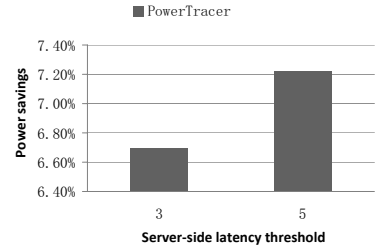}
\caption{The total system power savings of PowerTracer under two latency thresholds. }
\label{powertracer_powersavings_2factor}
\end{figure}

\begin{figure}[h]
\centering
\includegraphics[scale=0.60]{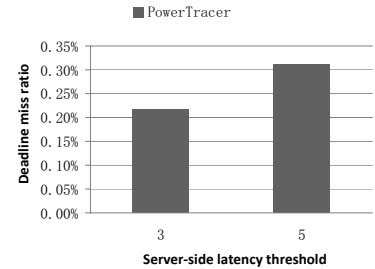}
\caption{The request deadline miss ratios of PowerTracer under two latency threshold. The deadline is set to 0.2 second.}
\label{powertracer_missratio_2factor}
\end{figure}

\begin{figure}[h]
\centering
\includegraphics[scale=0.60]{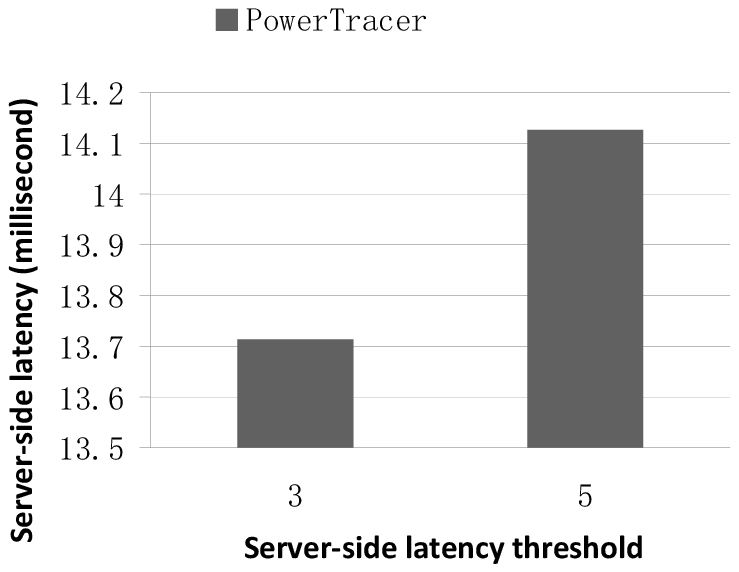}
\caption{The average server-side latencies of PowerTracer under two latency threshold.}
\label{powertracer_avgdelay_2factor}
\end{figure}

Fig. \ref{powertracer_powersavings_2factor} \textasciitilde{} Fig.\ref{powertracer_avgdelay_2factor} illustrate how the performance varies when the reference inputs in PowerTracer---the latency threshold varies. For this set of experiments, we set the number of clients to 500 and select three main patterns out of the top eight patterns. For the top eight patterns, $\overrightarrow{SL}$ is (0.15625, 9.207284, 13.67607, 11.8227, 28.14555, 22.5599, 22.7111, 22.50751) milliseconds. Fig. \ref{powertracer_powersavings_2factor} \textasciitilde{} Fig.\ref{powertracer_avgdelay_2factor} indicate that when the latency threshold is $5 \times \overrightarrow{SL}$, PowerTracer is only about 8\% better than that when the latency threshold is $3 \times \overrightarrow{SL}$ in terms of the power saving. Meanwhile, under these two configurations, their server-side latencies and their deadline miss ratios are close to each other.

\begin{figure}[h]
\centering
\includegraphics[scale=0.60]{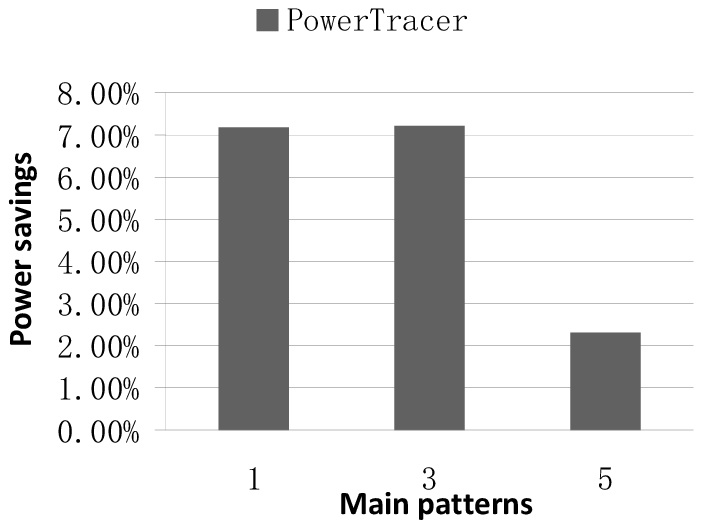}
\caption{A comparison of the total system power savings with different number of main patterns.}
\label{transition_powersavings_mainpatterns}
\end{figure}

\begin{figure}[h]
\centering
\includegraphics[scale=0.60]{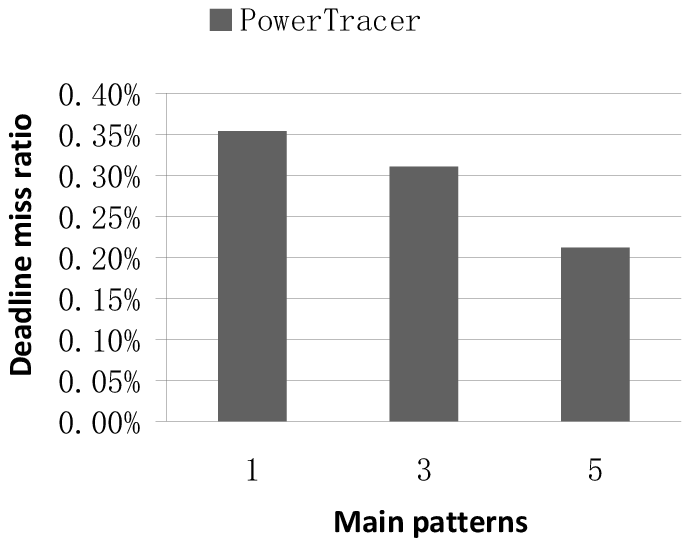}
\caption{A comparison of the request deadline miss ratios with different number of main patterns.}
\label{transition_missratio_mainpatterns}
\end{figure}

\begin{figure}[h]
\centering
\includegraphics[scale=0.60]{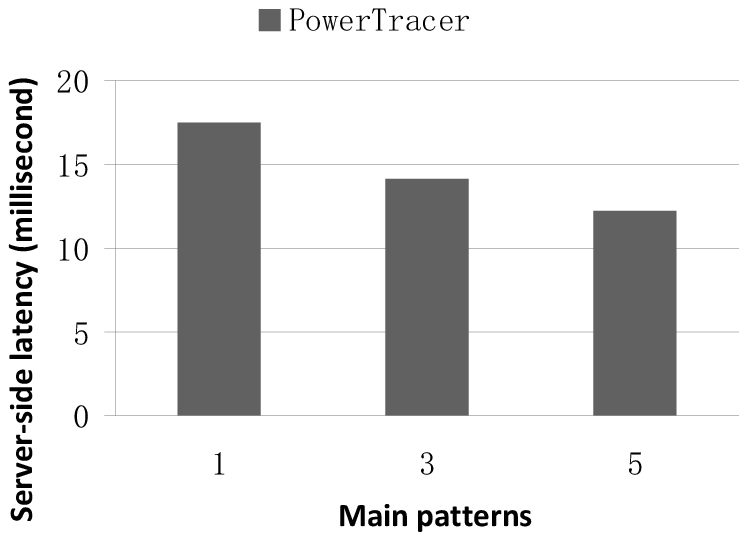}
\caption{A comparison of the average server-side latencies with different number of main patterns.}
\label{transition_avgdelay_mainpatterns}
\end{figure}

Fig.\ref{transition_powersavings_mainpatterns} \textasciitilde{} Fig.\ref{transition_avgdelay_mainpatterns} present the performance of PowerTracer when a different number of main patterns is chosen out of the top eight patterns. For these experiments, we set the number of clients to 500 and the latency threshold to $5 \times \overrightarrow{SL}$. Our results show that for the Read\_write mixed workload, PowerTracer achieves the optimal results,  i.e.,  the higher power saving, the lower miss ratio, and the lower server-side latency when we set the number of main patterns to three.


\subsection{Discussion}\label{discussion}
For two different workloads of RUBiS, the experiment results show that PowerTracer or PowerTracer\_NP outperforms its peer, indicating that request tracing can improve the accuracy of DVFS control.
At most of time, PowerTracer outperforms PowerTracer\_NP. This implies that monitoring the performance data of main patterns, instead of an average one, can also improve the accuracy of DVFS control. However, the optimal number of main patterns depends on different workloads. We also observe that setting a higher latency threshold under a certain limit in PowerTracer
improves power savings.

\textbf{Generality}. Our system is dependent on server-side latencies and service time per tier of the top $N$ patterns of a multi-tier service, which can be obtained with either black-box or white-box request tracing approaches. While PowerTracer is built on a basis of the black-box request tracing approach, our system can be built upon other white-box or black-box ones, like \cite{26Barham2004} \cite{27Koskinen2008} \cite{28Agarwala2007} \cite{29Chen2002} \cite{30Tak2009} \cite{Dapper}.

\textbf{Scalability}. In our previous work PreciseTracer \cite{scalable_tracing}, we have demonstrated how to improve system scalability through two mechanisms: tracing on demand and sampling. Besides, our experiments in \cite{scalable_tracing} show that PreciseTracer has fast responsiveness, and imposes negligible effects on the throughput and the average response time of services, like RUBiS. On a basis of those features, PowerTracer is promising in its system scalability.

\textbf{Potential for power saving}. So far, in most of commercial servers, CPU is the only energy proportional  component, and our work is also confined by this limitation. Barroso et al. \cite{25Barroso2007} showed that four components, including CPU, DRAM, disk, and network switches, are the main sources of power consumptions in data center. Therefore, we believe that our accurate DVFS control will play a more important role in  saving power consumption
when the concept of DVFS is extended to the other system components.

\section{Related Work} \label{related_work}
The related work can be classified into four categories: DVFS in server clusters, dynamic cluster reconfiguration, virtual machine based  server consolidation and power provisioning. 

\textbf{DVFS in server clusters.} The closest work  to this paper is \cite{19Horvath2007} by Horvath et al. They proposed a coordinated distributed DVS policy based on a feedback controller for three-tier web server systems. However, their work fails to provide accurate DVFS control for two reasons: first, the simple DVS algorithm uses CPU utilization as the indicator in determining which server's clock frequency should be scaled, while the optimized algorithm is difficult to be applied because of its complexity; second, the two algorithms  take the average server-side latency of all requests as the controlled variable, while our experiments show that massive requests have a number of different patterns.  In \cite{11Horvath2008},  Horvath et al. proposed a multi-mode energy management for multi-tier server clusters, which exploited DVS together with multiple sleep states. In \cite{12Horvath},  Horvath et al. invented a service prioritization scheme for multi-tier web server clusters, which assigned different priorities based on their performance requirements.  In \cite{24Chen2010},  Chen et al.  developed a simple metric called \emph{frequency gradient} that 
can predict the impact of changes in processor frequency upon the end-to-end transaction response times of multi-tier applications.

\textbf{Dynamic cluster reconfiguration.} K. Rajamani et al. \cite{request_distribution_2003} improved energy efficiency by powering down some servers when the desired quality of service can be met with fewer servers. M. Elnozahy et al. \cite{request_batch} used \emph{request batching} to conserve energy during periods of low workload intensity. Facing challenges in the context of connection servers, G. Chen et al. \cite{13Chen2008} designed a server provisioning algorithm to dynamically turn on a minimum number of servers, and a load dispatching algorithm to distribute load among the running machines. In integrating independently energy saving  policies, J. Heo et al. \cite{RTSS_2007}  presented a mechanism, called adaptation graph analysis, for identifying potential incompatibilities between composed adaptation policies.

\textbf{Virtual machine based server consolidation.} G. Dhiman et al \cite{vGreen} indicated that co-scheduling VMs with heterogeneous characteristics on the same physical node is beneficial from both energy efficiency and performance point of view. Y. Wang et al. \cite{virtual_batching} proposed Virtual Batching, a novel request batching solution for virtualized servers with primarily light workloads. Y. Wang et al. \cite{PARTIC} proposed a two-layer control architecture based on well-established control theory. X. Wang et al. \cite{co_co} proposed Co-Con, a cluster-level control architecture that coordinates individual power and performance control loops for virtualized server clusters. P. Padala et al. \cite{euro_sys_2007} developed an adaptive resource control system that dynamically adjusts the resource shares to individual tiers in order to meet application-level QoS goals . In their later work \cite{euro_sys_2009}, P. Padala et al. present AutoControl, a resource control system that automatically adapts to dynamic workload changes to achieve application
SLOs.

\textbf{Power provisioning.} Related work focuses on the performance optimization problem \emph{under power constraints}, while our work addresses the power optimization problem under performance constraints \cite{11Horvath2008}. P. Ranganathan et al. \cite{10RANGANATHAN2006} proposes power efficiencies at a larger scale by leveraging statistical properties of concurrent resource usage across a collection of systems (¡°ensemble¡±). C. Lefurgy et al. \cite{2Lefurgy2007} present a technique that controls the
peak power consumption of a high-density server . X. Wang et al. \cite{4Wang2008} \cite{TPDS_2010} propose a cluster-level power controller that shifts power among servers based on their performance needs, while controlling the total power of the cluster to be lower than a constraint. S. Pelley et al.\cite{Power_routing} develop mechanisms to better utilize installed power infrastructure. S. Govindan et al. \cite{statistical_profiling} explore a combination of statistical multiplexing techniques to improve the utilization
of the power hierarchy within a data center. R. Raghavendra et al. \cite{no_power_strugging} propose and validate a power management solution that coordinates different individual energy-saving approaches.  X. Fan et al. \cite{power_provisoning} present the aggregate power usage characteristics of large collections of servers (up to 15 thousand) for different classes of applications over a period of approximately six months, and modeling to attack data center-level power provisioning inefficiencies.

\section{Conclusion}\label{conclusion}
In this paper, we have proposed a novel request tracing approach  for cluster-level DVFS control. A request tracing tool can characterize major causal path patterns in serving different requests and capture server-side latency, especially service time of each tier in different patterns. The advantage of the request tracing approach is two-fold: first, it decreases the time cost of performance profiling experiments; second, it decreases the controller complexity so that we introduce a simpler feedback controller, which only relies on the single-node DVFS modulation at a time. Based on the request tracing approach, we have presented a hybrid DVFS control algorithm that combines an empirical performance model for fast modulation at different load levels and a simpler controller for adaption. We have developed a prototype of the proposed system, called PowerTracer, and conducted real experiments on a 3-tier platform to evaluate its performance. Our experimental results show that PowerTracer outperforms its peer \cite{19Horvath2007} in terms of power saving.


\begin{thebibliography}{99}

\bibitem{2Lefurgy2007}
C. Lefurgy, X. Wang, and M. Ware. Server-Level Power Control. Proceedings of the Fourth International Conference on Autonomic Computing, Washington, DC,USA, 2007. IEEE Computer Society.
\bibitem{4Wang2008}
X. Wang, M. Chen. Cluster-level feedback power control for performance optimization. HPCA 2008: 101-110.
\bibitem{6Sang2009}
B. Sang, J.-F. Zhan, G.-H. Tian, Decreasing log data of multi-tier services for effective request tracing, In Proc. International Conference on Dependable Systems and Networks (DSN'09), (2009) H22-H23.
\bibitem{10RANGANATHAN2006}
P. Ranganathan, P. Leech, D. Irwin, and J. Chase,  2006. Ensemble-level Power Management for Dense Blade Servers. SIGARCH Comput. Archit. News 34, 2 (May. 2006), 66-77.
\bibitem{11Horvath2008}
T. Horvath and K. Skadron. Multi-mode energy management for multi-tier server clusters. International Conference on Parallel Architectures and Compilation Techniques (PACT), 2008.
\bibitem{12Horvath}
T. Horvath, K. Skadron and T. Abdelzaher. Enhancing energy efficiency in multi-tier web server clusters via prioritization.  In IPDPS '07, pages 1--6. IEEE Computer Society, 2007.
\bibitem{13Chen2008}
G. Chen, W. He, J. Liu, S. Nath, L. Rigas, L. Xiao, and F. Zhao. Energy-Aware Server Provisioning and Load dispatching for Connection-Intensive Internet Services. NSDI, 2008.
\bibitem{14Zhang2009}
Z. Zhang, J. Zhan, Y. Li, L. Wang, D. Meng, B. Sang. Precise Request Tracing and Performance Debugging for Multi-tier Service of Black Boxes. in Proceedings of DSN 2009.
\bibitem{scalable_tracing}
B. Sang, J. Zhan, Z. Zhang, L. Wang, D. Xu, Y. Huang and D. Meng, Precise, scalable and online request tracing for multi-tier services of black boxes, Technical Report. http://arxiv4.library.cornell.edu/abs/1007.4057v1.
\bibitem{15rubis}
http://rubis.objectweb.org.
\bibitem{17Chase2001}
J. Chase, D. Anderson, P. Thakar, A. Vahdat, and R. Doyle. Managing energy and server resources in hosting centers. Proceedings of the 18th Symposium on Operating Systems Principles (SOSP), Oct. 2001.
\bibitem{18HEATH2005}
HEATH, T., DINIZ, B., ET AL. Energy conservation in heterogeneous server clusters. In Proceedings of PPoPP 05.
\bibitem{19Horvath2007}
T. Horvath, T. Abdelzaher, K. Skadron, and X. Liu. Dynamic voltage scaling in multi-tier web servers with end-to-end delay control. IEEE Transactions on Computers, vol. 56, no. 4, pp. 444-458, 2007.
\bibitem{22Hartigan1979}
J. A. Hartigan, M. A. Wong, A k-means cluster algorithm, Applied Statistics, 28, 1979, pp. 100-108.
\bibitem{23Urgaonkar2005}
Urgaonkar, B., Pacifici, G., Shenoy, P., Spreitzer, M., and Tantawi, A. 2005. An analytical model for multi-tier internet services and its applications. SIGMETRICS Perform. Eval. Rev. 33, 1 (Jun. 2005), 
\bibitem{24Chen2010}
Chen, S., Joshi, K. R., Hiltunen, M. A., Schlichting, R. D., and Sanders, W. H. 2010. Blackbox prediction of the impact of DVFS on end-to-end performance of multitier systems. SIGMETRICS Perform. Eval. Rev. 37, 4 (Mar. 2010), 59-63.
\bibitem{25Barroso2007}
Barroso, L. A. and H?lzle, U. 2007. The Case for Energy-Proportional Computing. Computer 40, 12 (Dec. 2007), 33-37.
\bibitem{26Barham2004}
P. Barham, A. Donnelly, R. Isaacs, R. Mortier, Using Magpie for Request Extraction and Workload Modeling, In Proc. of the 6th USENIX OSDI, (2004) 259-272.
\bibitem{27Koskinen2008}
E. Koskinen, J. Jannotti, BorderPatrol: Isolating Events for Precise Black-box Tracing, In Proc. of 3rd ACM SIGOPS/European Conference on Computer Systems, (2008).
\bibitem{28Agarwala2007}
S. Agarwala, F. Alegre, K. Schwan, J. Mehalingham, E2EProf: Automated End-to-End Performance Management for Enterprise Systems, In Proc. International Conference on Dependable Systems and Networks (DSN'07), (2007) 749-758.
\bibitem{29Chen2002}
M.-Y. Chen, E. Kiciman, E. Fratkin, A. Fox, E. Brewer, Pinpoint: Problems Detection in Large, Dynamic Internet Service, In Proc. of International Conference on Dependable Systems and Networks (DSN'02), (2002) 23-26.
\bibitem{30Tak2009}
B.-C. Tak, C. Tang, C. Zhang, S. Govindan, B. Urgaonkar, vPath: Precise Discovery of Request Processing Paths from Black-Box Observations of Thread and Network Activities, USENIX'09, (2009).
\bibitem{31ondemand}
http://xlife.zuavra.net/index.php/70/\#cpufreq\textbackslash{}-governors
\bibitem{virtual_batching}
Y. Wang, R. Deaver and X. Wang, Virtual Batching: Request Batching for Energy Conservation in Virtualized Servers ", In Proceedings of IWQoS 2010.
\bibitem{request_batch}
M. Elnozahy, M. Kistler, and R. Rajamony,  2003. Energy conservation policies for web servers. In Proceedings of USIT 04. USENIX Association.
\bibitem{RTSS_2007}
J. Heo, D. Henriksson, X. Liu, T. F. Abdelzaher: Integrating Adaptive Components: An Emerging Challenge in Performance-Adaptive Systems and a Server Farm Case-Study. RTSS 2007: 227-238.
\bibitem{TPDS_2010}
X. Wang, M. Chen, and X. Fu, MIMO Power Control for High-Density Servers in an Enclosure, IEEE Transactions on Parallel and Distributed Systems, 21(1): 1-15, January 2010.
\bibitem{PARTIC}
Y. Wang, X. Wang, M. Chen, and X. Zhu, PARTIC: Power-Aware Response Time Control for Virtualized Web Servers, IEEE Transactions on Parallel and Distributed Systems, 1(1): 1-15, January 2010.
\bibitem{Power_routing}
 S. Pelley,D. Meisner, P. Zandevakili, T. F.Wenisch, J. and Underwood,  2010. Power routing: dynamic power provisioning in the data center. SIGPLAN Not. 45, 3 (Mar. 2010), 231-242.
\bibitem{energy_efficient_2002}
 E. Elnozahy, M. Kistler, and R. Rajamony. Energy-efficient server clusters. In Proc. Workshop on Power-Aware Computing Systems, Feb. 2002.
 \bibitem{request_distribution_2003}
 K. Rajamani and C. Lefurgy. On evaluating request-distribution schemes for saving energy in server clusters. In Proc. IEEE International Symposium on Performance Analysis of Systems and Software, pages 111-122, 2003.
\bibitem{energy_efficient_real_time_2006}
C. Rusu, A. Ferreira, C. Scordino, and A. Watson. Energy-efficient real-time heterogeneous server clusters. In Proc. 12th IEEE Real-Time and Embedded Technology and Applications Symposium, pages 418-428, 2006.
\bibitem{power_provisoning}
X. Fan, W.-D. Weber, and L. A. Barroso. Power provisioning for a warehouse-sized computer. In Proc.34th Annual ACM/IEEE International Symposium on Computer Architecture, pages 13-23, 2007.
\bibitem{no_power_strugging}
R. Raghavendra, P. Ranganathan, V. Talwar, Z. Wang, and X. Zhu. No "power" struggles: coordinated multi-level power management for the data center. SIGARCH Comput. Archit. News, 36(1):48-59, 2008.
\bibitem{statistical_profiling}
Govindan, S., Choi, J., Urgaonkar, B., Sivasubramaniam, A., and Baldini, A. 2009. Statistical profiling-based techniques for effective power provisioning in data centers. In Proceedings of EuroSys '09. ACM, New York, NY, 317-330.
\bibitem{vGreen}
Dhiman, G., Marchetti, G., and Rosing, T. 2009. vGreen: a system for energy efficient computing in virtualized environments. In Proceedings of ISLPED '09. ACM, New York, NY, 243-248.
\bibitem{RTSS}
Y. Wang, X. Wang, M. Chen, and  X. Zhu, 2008. Power-Efficient Response Time Guarantees for Virtualized Enterprise Servers. In Proceedings of RTSS 08. 303-312.
\bibitem{co_co}
X. Wang and Y. Wang, Coordinating Power Control and Performance Management for Virtualized Server Clusters, IEEE Transactions on Parallel and Distributed Systems, 2010.
\bibitem{power_nap}
D. Meisner, B. T. Gold, and T. F. Wenisch,  2009. PowerNap: eliminating server idle power. In Proceeding of ASPLOS '09. ACM, New York, NY, 205-216.
\bibitem{control_theory}
X. Zhu, M. Uysal, Z. Wang, S. Singhal, A. Merchant, P. Padala, and K. Shin, 2009. What does control theory bring to systems research?. SIGOPS Oper. Syst. Rev. 43, 1 (Jan. 2009), 62-69.
\bibitem{non_proportional}
N. Tolia, Z. Wang, M. Marwah, C. Bash, P. Ranganathan, and X. Zhu. Delivering Energy Proportionality with Non Energy-Proportional Systems¡ªOptimizing the Ensemble. In USENIX HotPower, 2008.
\bibitem{euro_sys_2007}
P. Padala, K. G. Shin, X. Zhu,  M. Uysal, Z. Wang, S. Singhal,  A. Merchant, and K. Salem,  2007. Adaptive control of virtualized resources in utility computing environments. SIGOPS Oper. Syst. Rev. 41, 3 (Jun. 2007), 289-302.
\bibitem{euro_sys_2009}
P. Padala, K. Hou, K. G. Shin, X. Zhu, M. Uysal, Z. Wang,  S. Singhal, and  A. Merchant. 2009. Automated control of multiple virtualized resources. In Proceedings of the 4th ACM European Conference on Computer Systems (Nuremberg, Germany, April 01 - 03, 2009). EuroSys '09. ACM, New York, NY, 13-26.

\bibitem{HPC_placement_2009}
A. Verma, P. Ahuja,  A. and Neogi, 2008. Power-aware dynamic placement of HPC applications. In Proceedings ICS '08.

\bibitem{SLA_decomposion}
Y. Chen, S. Iyer, X. Liu, D. Milojicic,  A. and Sahai, 2007. SLA Decomposition: Translating Service Level Objectives to System Level Thresholds. In Proceedings of ICAC 07.


\bibitem{Dapper}
B. H. Sigelman, L. A. Barroso, M. Burrows, P. Stephenson, M. Plakal, D.Beaver, S.Jaspan, C. Shanbhag, \emph{Dapper, a Large-Scale Distributed Systems Tracing Infrastructure}, Google Technical Report dapper-2010-1, April 2010.

\end{thebibliography}
\end{document}